# Interactive Supercomputing on 40,000 Cores for Machine Learning and Data Analysis

Albert Reuther, Jeremy Kepner, Chansup Byun, Siddharth Samsi, William Arcand, David Bestor, Bill Bergeron,
Vijay Gadepally, Michael Houle, Matthew Hubbell, Michael Jones, Anna Klein, Lauren Milechin, Julia Mullen,
Andrew Prout, Antonio Rosa, Charles Yee, Peter Michaleas
Massachusetts Institute of Technology Lincoln Laboratory Supercomputing Center
Lexington, MA, USA

*Abstract*—Interactive massively parallel computations are critical for machine learning and data analysis. These computations are a staple of the MIT Lincoln Laboratory Supercomputing Center (LLSC) and has required the LLSC to develop unique interactive supercomputing capabilities. Scaling interactive machine learning frameworks, such as TensorFlow, and data analysis environments, such as MATLAB/Octave, to tens of thousands of cores presents many technical challenges – in particular, rapidly dispatching many tasks through a scheduler, such as Slurm, and starting many instances of applications with thousands of dependencies. Careful tuning of launches and prepositioning of applications overcome these challenges and allow the launching of thousands of tasks in seconds on a 40,000-core supercomputer. Specifically, this work demonstrates launching 32,000 TensorFlow processes in 4 seconds and launching 262,000 Octave processes in 40 seconds. These capabilities allow researchers to rapidly explore novel machine learning architecture and data analysis algorithms.

## I. INTRODUCTION

Interactive supercomputing has been an ongoing focal point of high performance computing (HPC) at Lincoln Laboratory [Reuther 2004]. Since its inception, users have connected their desktops and laptops to Lincoln's interactive supercomputer and been able to launch parallel pMatlab jobs from their desktop/laptop integrated developer environment (IDE) [Reuther 2005].

This system architecture has evolved into the MIT SuperCloud, a fusion of the four large computing ecosystems – supercomputing, enterprise computing, big data and, traditional databases – into a coherent, unified platform that enables rapid prototyping capabilities across all four computing ecosystems. The MIT SuperCloud has spurred the development of a number of cross-ecosystem innovations in high performance databases [Byun 2012], [Kepner 2014a], database management [Prout 2015], data protection [Kepner 2014b], database federation [Kepner 2013], [Gadepally 2015], data analytics [Kepner 2012] and system monitoring [Hubbell 2015].

This capability has grown in many dimensions. The MIT Lincoln Laboratory Supercomputing Center (LLSC) provides interactive supercomputing to thousands of users at MIT Lincoln Laboratory and at the MIT Beaver Works Center for Engaging Supercomputing. LLSC not only continues to support parallel MATLAB and Octave jobs, but also jobs in Python [Van Rossum 2007], Julia [Bezanson 2017], R [Ihaka 1996], Tensorflow [Abadi 2016], PyTorch [Paszke 2017], and Caffe [Jia 2014] along with parallel C, C++, Fortran, and Java applications with various flavors of message passing interface (MPI). Furthermore, the TX-Green flagship system now has nearly 60,000 cores available for users' parallel jobs. The most significant jump in core count was the addition of 648 Intel Xeon Phi 64-core nodes [Byun 2017, Cichon 2016], each of which has 64 compute cores in a single processor socket laid out in a mesh configuration [Jeffers 2016]. This equals 41,472 total cores across the 648 compute nodes, all connected by a non-blocking 10-Gigabit Ethernet network and a non-blocking Intel OmniPath low-latency network.

Scaling immediate interactive launches to such a large number of cores was a significant challenge; this paper discusses the technical experimentation and engineering involved in scaling the interactive parallel launching capability of TX-Green to the scale of 40,000 core jobs. In the Section II, we review the background of interactive supercomputing, discuss the components of a supercomputing scheduler, review the results of a previous study comparing state-of-the-art HPC schedulers and resource managers. Section III details the experimentation and steps taken to enable interactive supercomputing launches to the scale of 40,000 core jobs, while

This material is based upon work supported by the Assistant Secretary of Defense for Research and Engineering under Air Force Contract No. FA8721-05-C-0002 and/or FA8702-15-D-0001. Any opinions, findings, conclusions or recommendations expressed in this material are those of the author(s) and do not necessarily reflect the views of the Assistant Secretary of Defense for Research and Engineering.

Section IV discusses the scaling launch time results on the 64-core Xeon Phi compute nodes. Finally, the paper is summarized in Section V.

## II. INTERACTIVE SUPERCOMPUTING LAUNCH

Whether on a laptop or smartphone, interactivity is inherent in our daily interactions with computers since that computer is dedicated exclusively to ourselves when we are interacting with the device. However, supercomputers are almost always a shared set of resources. Traditionally, supercomputer jobs were submitted to a job queue, from which the scheduler chose the optimal job to execute next when resources became available. This scheduling technique is called batch scheduling, and it introduces latency between job submission and job execution as depicted in Figure 1 [Reuther 2007]. However, one component of interactive supercomputing is enabling very fast parallel on-demand (immediate) job launches, while the other main component is supporting parallel high productivity software packages including MATLAB/Octave, Python, Julia, and R along with domain specific packages like Tensorflow, Caffe, and PyTorch. In this paper, we focus on the job launches to enable the efficient use of such high productivity software packages. Such interactive launches are also depicted in Figure 1, and this workflow does not have time spent in the pending state.

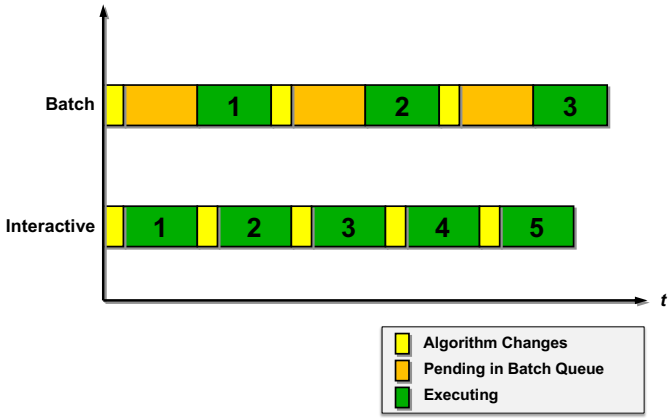

Figure 1. Batch versus interactive launch cycles.

On-demand (immediate) parallel job launches and interactive environments empower rapid prototyping, debugging, testing of code, analysis, and visualization on a wide variety of types of jobs. Faster launches lead to faster development turns and greater development turns lead to greater efficiency and productivity. The challenges lie in enabling interactive, on-demand parallel jobs as depicted in Figure 2. At one extreme, all jobs are scheduled as batch jobs which can incur high latency before execution, while at the other extreme all jobs are scheduled immediately, which can cause scheduler flooding. Most supercomputing centers use batch queuing with reservations, which allow users to reserve a set of resources sometime in the future for a window of interactive computing. For the LLSC, we have chosen the route of interactive, immediate launches with user resource limits. This enables immediate interactive jobs, while avoiding scheduler flooding.

|  | All Batch | Batch with Reservations | Interactive with Limits | All Immediate Scheduling |
|---|---|---|---|---|
| Pro | Jobs launches optimize system utilization | Jobs launches maximize system utilization; Reservations specify timeslots to interact with job | Launch most jobs when submitted; Limits promotes fairness; Accommodates both interactive and batch jobs | Run all jobs immediately |
| Con | Often high latency before execution | Often high latency before execution | Lower system utilization | Frequent scheduler flooding; Lower system utilization |

Figure 2. Interactive scheduler tradeoffs.

To better understand how we have implemented our interactive scheduler, we must discuss the various components of the job scheduler. At its simplest level, job schedulers are responsible for matching and executing compute jobs from different users on computational resources. The users and their jobs will have different resource requirements and priorities. Similarly, the computational resources have different resource availabilities and capabilities, and they must be managed in such a way that they are best utilized, given the mix of jobs that need to be executed.

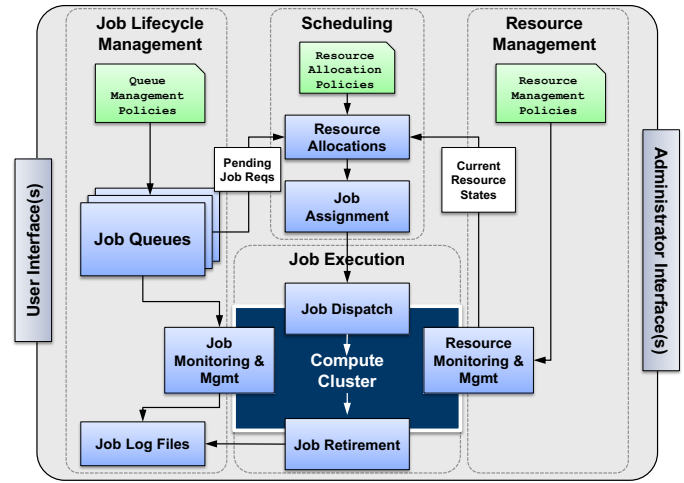

Figure 3. Scheduler architecture.

A cluster job scheduler has four key operational tasks: job lifecycle management, resource management, scheduling, and job execution, as shown in Figure 3. The job lifecycle management task receives jobs from users through the user interface and places them in one of the job queues to wait for execution (regardless of whether jobs are scheduled and executed on demand or batch queued). Various resources for the job including memory, licenses, and accelerators (such as GPUs) are requested through the user interface by the user. The job lifecycle management task is also responsible for prioritizing and sorting candidate jobs for execution by using the queue management policies. The scheduling task periodically requests a prioritized list of candidate queued jobs and determines whether resources are available to execute one or more of the jobs. The scheduler receives the state of all the resources from the resource management task, which in turn is receiving resource state and availability information from the compute nodes. The scheduling task allocates resources (usually one or more job slots on compute nodes) and assigns the job to the resource(s) if adequate resources are available to execute each job. The job execution task is responsible for dispatching/launching the job on the resources. Upon the completion of each job, the job execution task manages the closing down of the job and reporting the statistics for the job to the job lifecycle management task, which records it in logs.

In recent studies [Reuther 2016, Reuther 2018], we conducted a detailed comparison of traditional supercomputing schedulers and Big Data schedulers. Two of the most important takeaways from this comparison were:

1. The traditional supercomputing schedulers including Slurm [Yoo 2003], LSF [Zhou 1993], and GridEngine [Slapnicar 2001] were capable of launching synchronously parallel (MPI-style) jobs as well as loosely parallel job arrays. Big Data schedulers including Mesos, Apache YARN [Vavilapalli 2013], and the open-source Kubernetes project [Hindman 2011] supported only parallel job arrays.

2. Several schedulers including Slurm, Mesos, and Kubernetes were designed to handle 100,000+ jobs, both in its queues and executing on compute nodes.

Since Slurm supports both synchronously parallel jobs and job arrays and scaled to managing 100,000+ jobs, it substantiated the continued use of Slurm as the job scheduler for LLSC systems.

III. LAUNCHING 40,000 CORE JOBS

Recently, LLSC upgraded its flagship system with 648 Intel Xeon Phi compute nodes. Each node has a 64-core Intel Xeon Phi 7210 processor, for a total of 41,472 cores, along with 192 GB RAM, 16 GB of on-package MCDRAM configured in 'flat' mode, local storage, 10-GigE network interface, and an OmniPath application network interface each. The Lustre [Braam 2003] central storage system uses a 10 petabyte Seagate ClusterStor CS9000 storage array that is directly connected to the core switch. As with all of the LLSC systems, enabling interactive jobs was a top priority. However, the first attempts at launching interactive MATLAB/Octave jobs through slurm onto 40,000 processors resulted in 30- to 60-minute launch times; these launch times were a hindrance to any interactivity with the jobs.

To enable truly interactive launches, a number of experiments and engineering trade-offs were explored. First, we investigated how fast launches could be enabled. We started by allocating a block of nodes through Slurm with the salloc command, feeding the node list into pMatlab [Kepner 2009], and using a hierarchical secure shell (`ssh`) process spawning mechanism to launch a large set of interactive processes. This gave us a baseline for how fast we could expect to launch 10,000+ core jobs – launches of less than a minute should be possible. We went on to explore the use of job arrays and synchronously parallel launches, which each had their trade-offs. Synchronously parallel jobs using `srun` enabled the fastest launches, but the resources for a job remained allocated until all of the computational processes completed. Conversely, each job array process relinquishes its resources as soon as it finishes its work. Launch times were similar. We also experimented with various queue evaluation periodicities and job queue evaluation depth values to find the most effective combination.

To further speed up launches, we decided to allocate whole compute nodes and launch a single scheduler-issued launcher process per compute node. This launcher process subsequently spawns and backgrounds each of the application processes that are to be launched on its compute node.

We made several improvements in tuning the launching of applications themselves. First, we copied the entire installations of five MATLAB versions, two Octave versions, and five versions of Anaconda Python including TensorFlow, Caffe, and PyTorch onto the local hard drive of every compute node. This reduced the latency of loading thousands of instances from a central file system and across the 10Gig-E network. We also used the timing flag with MATLAB to record what segments of MATLAB startups used the most time and reduced its launch time further. This prompted us to also create a MATLAB-lite version which loaded only the base MATLAB toolboxes and did not include the internal Java invocation. With all of these improvements, we met the interactive launching goals that we had set out to achieve.

## IV. PERFORMANCE RESULTS

Training machine learning models requires high level programming environments for building the models and rapid interaction with the analyst to converge on the best training parameters. Standard approaches take minutes to hours to launch models on thousands of cores. However, with the improvements we discussed in the previous section, we are able to launch hundreds of machine learning models in a matter of seconds.

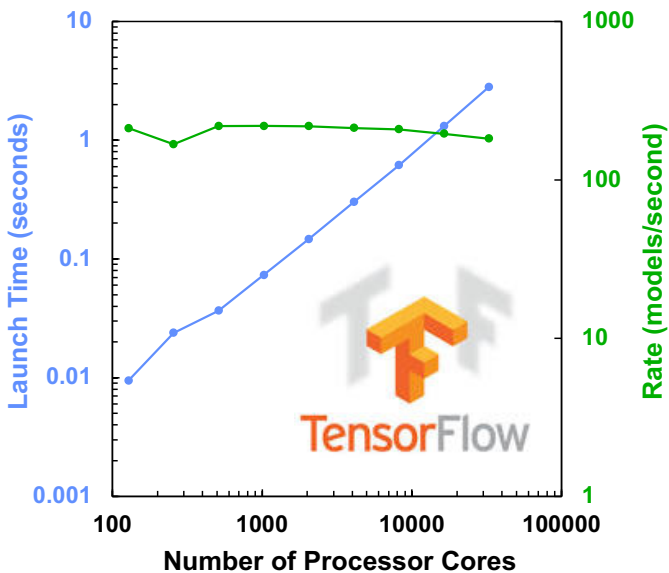

Figure 4: Tensorflow launch scaling results

TensorFlow is one of the leading deep neural network model frameworks available today. TensorFlow is supported by Google, and it provides a productive Python API for generating and training deep neural networks [Abadi 2016]. Figure 4 is a log-log plot scaling up the number of processor cores on the x-axis versus the launch time on the y-axis. We have achieved launch times of less than 5 seconds for 32,000+ cores (512 64-core Xeon nodes). In other words, we are able to launch 512 TensorFlow models simultaneously. This enables very rapid trade-off analyses of neural network batch size, convergence rates, input set randomization, etc. for a truly interactive machine learning experience.

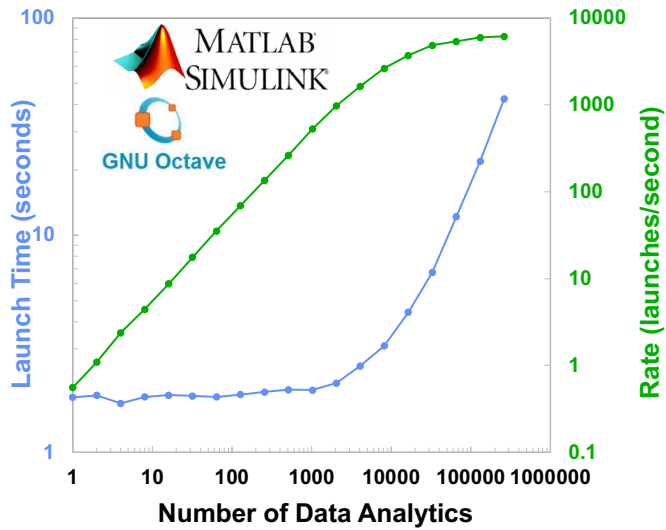

Figure 5: MATLAB/Octave launch scaling results.

Many researchers at MIT frequently use MATLAB and Octave for rapid prototyping, algorithm development, and data analysis. These activities require rapid interaction and fast turnarounds to make significant progress and convergence to a solution. We achieved similar launching results with pMatlab and parallel Octave jobs. Figure 5 is a log-log plot scaling up the number of processor cores on the x-axis versus the launch time on the y-axis. We have achieved launch times of less than 10 seconds for launching 32,000+ MATLAB/Octave jobs (512 64-core Xeon nodes) launching one MATLAB/Octave process per core. Furthermore, we have achieved parallel launches of 260,000+ MATLAB/Octave process launches in under 40 seconds. Each of the cores on a Xeon Phi processor has four hyperthreads, and this parallel launch involves launching 512 MATLAB/Octave processes per Xeon Phi processor, two for each hyperthread.
4

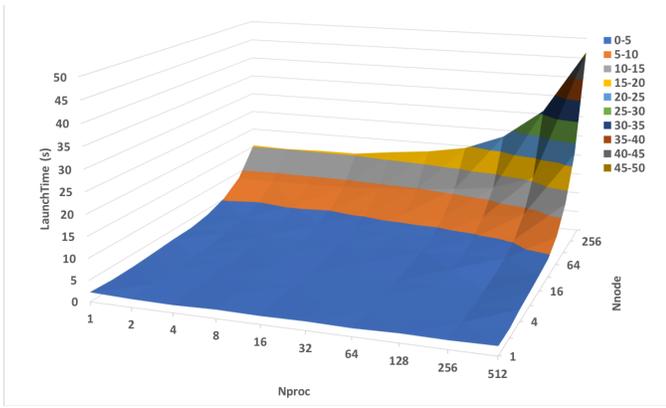

**Figure 6:** Launch times in seconds of paralllel MATLAB/Octave jobs over Nnode nodes and Nproc MATLAB/Octave processes per node.

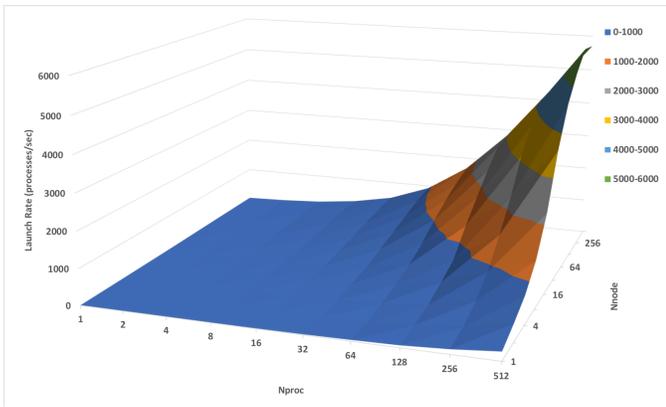

**Figure 7:** Launch rates of paralllel MATLAB/Octave jobs over Nnode nodes and Nproc MATLAB/Octave processes per node.

We have further measured the launch time and launch rate for the parallel MATLAB/Octave jobs, varying the number of nodes from 1 to 512 in powers of 2, and varying the number of processes per node from 1 to 512 in powers of 2. The launch time results are show in Figure 6; launch times remain under 10 seconds for all but the largest number of nodes onto which the processes were launched. Further, launch times are under 20 seconds but for the very largest node numbers and processes per node. Figure 7 displays the launch rates in process launches per second. This plot shows that the scheduler and the local launchers can sustain launch rates of 6,000 processes per second. We have found that Slurm handles the many parallel launches onto each of the nodes quite well. Our two-tiered launching mechanism is very effective on manycore processors such as the Intel Xeon Phi. We have determined that the rise in launch time for high node counts and processes per node arises from backpressure from our low-latency, high-bandwidth Lustre central file system, which serves a few files to each of the launching processes. However, serving a few files to each process when there are many processes does add up.

## V. SUMMARY

High performance launch at scale is a generally enabling capability of interactive supercomputing. It allows the processing of larger sets of sensor data, the creation of higher-fidelity simulations, and the development new algorithms for space observation, robotic vehicles, communications, cyber security, machine learning, sensor processing, electronic devices, bioinformatics, and air traffic control. In this paper, we have discussed the technical experimentation and engineering involved in scaling the interactive parallel launching capability of TX-Green to the scale of 40,000 core jobs. The applications for which we shared results are the TensorFlow machine learning framework and the MATLAB/Octave rapid prototyping language and environment. These launching capabilities enable very large Monte Carlo and parameter trade-off analyses using these very familiar frameworks and programming environments.

ACKNOWLEDGMENTS

The authors wish to acknowledge the following individuals for their contributions: Bob Bond, Alan Edelman, Chris Hill, Charles Leiserson, Dave Martinez, and Paul Monticciolo.

REFERENCES

[Abadi 2016] M. Abadi, P. Barham, J. Chen, Z. Chen, A. Davis, J. Dean, M. Devin, S. Ghemawat, G. Irving, M. Isard, M. Kudlur, J. Levenberg, R. Monga, S. Moore, D. Murray, B. Steiner, P. Tucker, V. Vasudevan, P. Warden, M. Wicke, Y. Yu, and X. Zheng, "TensorFlow: A System for Large-Scale Machine Learning," 12th USENIX Symposium on Operating System Design and Implementation (OSDI), Savannah, GA, 2016.

[Bezanson 2017] J. Bezanson, A. Edelman, S. Karpinski and V. B. Shah, "Julia: A Fresh Approach to Numerical Computing," *SIAM Review,* vol. 59, pp. 65-98, 2017.

[Braam 2003] P. J. Braam, et.al., "The Lustre Storage Architecture, Cluster File Systems, Inc., October 2003.

[Byun 2012] C. Byun, W. Arcand, D. Bestor, B. Bergeron, M. Hubbell, J. Kepner, A. McCabe, P. Michaleas, J. Mullen, D. O'Gwynn, A. Prout, A. Reuther, A. Rosa, and C. Yee, "Driving Big Data with Big Compute." IEEE *High Performance Extreme Computing Conference (HPEC)*, Waltham, MA, September 10-12, 2012.

[Byun 2017] C. Byun, J. Kepner, W. Arcand, D. Bestor, B. Bergeron, V. Gadepally, M. Houle, M. Hubbell, M. Jones, A. Klein, P. Michaleas, L. Milechin, J. Mullen, A. Prout, A. Rosa, S. Samsi, C. Yee, A. Reuther, "Benchmarking Data Analysis and Machine Learning Applications on the Intel KNL Many-Core Processor," *IEEE High Performance Extreme Computing (HPEC) Conference*, Waltham, MA, September 12-14, 2017.

[Cichon 2016] M. Cichon, "Lincoln Laboratory's Supercomputing System Ranked Most Powerful in New England," MIT Lincoln Laboratory News, November 2016. URL: https://www.ll.mit.edu//news/LLSC-supercomputing-system.html

[Gadepally 2015] V. Gadepally, J. Kepner, W. Arcand, D. Bestor, B. Bergeron, C. Byun, L. Edwards, M. Hubbell, P. Michaleas, J. Mullen, A. Prout, A. Rosa, C. Yee, A. Reuther, "D4M: Bringing Associative Arrays to Database Engines," IEEE High Performance Extreme Computing Conference (HPEC), Waltham, MA September 15-17, 2015.

[Hindman 2011] B. Hindman, A. Konwinski, M. Zaharia, A. Ghodsi, A. D. Joseph, R. H. Katz, S. Shenker, and I. Stoica. "Mesos: A Platform for Fine-Grained Resource Sharing in the Data Center," NSDI, vol. 11, pp. 22-22. 2011.

[Hubbell 2015] M. Hubbell, A. Moran, W. Arcand, D. Bestor, B. Bergeron, C. Byun, V. Gadepally, P. Michaleas, J. Mullen, A. Prout, A. Reuther, A. Rosa, C. Yee, J. Kepner, "Big Data Strategies for Data Center Infrastructure Management Using a 3D Gaming Platform," IEEE High Performance Extreme Computing Conference (HPEC), Waltham, MA, September 15-17, 2015.

[Ihaka 1996] R. Ihaka and R. Gentleman, R: a Language for Data Analysis and Graphics," *Journal of Computational and Graphical Statistics*, vol. *5*, no. 3 , pp.299-314, 1996.

[Jeffers 2016] J. Jeffers, J. Reinders, and A. Sodani, *Intel Xeon Phi Processor High Performance Programming: Knights Landing Edition,* Second Edition, Elsevier, 2016.

[Jia 2014] Y. Jia, E. Shelhamer, J. Donahue, S. Karayev, J. Long, R. Girshick, S. Guadarrama, and T. Darrell, "Caffe: Convolutional Architecture for Fast Feature Embedding," *Proceedings of ACM Multimedia,* pp. 675-678, 2014.

[Kepner 2009] J. Kepner, *Parallel Matlab for Multicore and Multinode Computers*, SIAM Press, 2009.

[Kepner 2012] J. Kepner, W. Arcand, W. Bergeron, N. Bliss, R. Bond, C. Byun, G. Condon, K. Gregson, M. Hubbell, J. Kurz, A. McCabe, P. Michaleas, A. Prout, A. Reuther, A. Rosa and C. Yee, "Dynamic Distributed Dimensional Data Model (D4M) Database and Computation System," *IEEE International Conference on Acoustics, Speech and Signal Processing (ICASSP)*, pages 5349–5352, 2012.

[Kepner 2013] J. Kepner, C. Anderson, W. Arcand, D. Bestor, B. Bergeron, C. Byun, M. Hubbell, P. Michaleas, J. Mullen, D. O'Gwynn, A. Prout, A. Reuther, A. Rosa, and C. Yee, "D4M 2.0 Schema: A General Purpose High Performance Schema for the Accumulo Database," IEEE High Performance Extreme Computing (HPEC) Conference, Waltham, MA, Sep 10-12, 2013.

[Kepner 2014a] J. Kepner, W. Arcand, D. Bestor, B. Bergeron, C. Byun, V. Gadepally, M. Hubbell, P. Michaleas, J. Mullen, A. Prout, A. Reuther, A. Rosa, and C. Yee, "Achieving 100,000,000 Database Inserts per Second Using Accumulo and D4M," IEEE High Performance Extreme Computing Conference (HPEC), Waltham, MA, September 9-11, 2014.

[Kepner 2014b] J. Kepner, V. Gadepally, P. Michaleas, N. Schear, M. Varia, A. Yerukhimovich, and R. K. Cunningham, "Computing on Masked Data: A High Performance Method for Improving Big Data Veracity," IEEE High Performance Extreme Computing Conference (HPEC), Waltham, MA, September 9-11, 2014.

[Paszke 2017] A. Paszke, S. Gross, S. Chintala, G. Chanan, E. Yang, Z. DeVito, Z. Lin, A. Desmaison, L. Antiga, A. Lerer, "Automatic Differentiation in PyTorch," NIPS-W, 2017.

[Prout 2015] A. Prout, J. Kepner, P. Michaleas, W. Arcand, D. Bestor, B. Bergeron, C. Byun, L. Edwards, V. Gadepally, M. Hubbell, J. Mullen, A. Rosa, C. Yee, A. Reuther, "Enabling On-Demand Database Computing with MIT SuperCloud Database Management System," IEEE High Performance Extreme Computing Conference (HPEC), Waltham, MA, September 15-17, 2015.

[Reuther 2004] A.I. Reuther, T. Currie, J. Kepner, H.G. Kim, A. McCabe, M.P. Moore, N. Travinin, "On-Demand Grid Computing Using gridMatlab and pMatlab," *Proceedings of the High Performance Computing Modernization Office Users Group Conference 2004*, Williamsburg, VA, 8 June 8, 2004.

[Reuther 2005] A. Reuther, T. Currie, J. Kepner, H. Kim, A. McCabe, M. Moore and N. Travinin, "Technology Requirements for Supporting On-Demand Interactive Grid Commputing," *Proceedings of the DoD High Performance Computing Modernization Program (HPCMP) Users Group Conference (UGC)*, Nashville, TN, June 27-30, 2005.

[Reuther 2007] A. Reuther, J. Kepner, A. McCabe, J. Mullen, N.T. Bliss, and H. Kim, "Technical Challenges of Supporting Interactive HPC," *Proceedings of the High Performance Computing Modernization Program (HPCMP) Users Group Conference (UGC)*, Pittsburgh, PA, June 18-22 2007.

[Reuther 2013] A. Reuther, J. Kepner, W. Arcand, D. Bestor, W. Bergeron, C. Byun, M. Hubbell, P. Michaleas, J. Mullen, A. Prout, and A. Rosa, "LLSuperCloud: Sharing HPC Systems for Diverse Rapid Prototyping," IEEE High Performance Extreme Computing (HPEC) Conference, Waltham, MA, Sep 10-12, 2013.

[Reuther 2016] A. Reuther, C. Byun, W. Arcand, D. Bestor, B. Bergeron, M. Hubbell, M. Jones, P. Michaleas, A. Prout, A. Rosa,. J. Kepner, "Scheduler Technologies in Support of High Performance Data Analysis," *IEEE High Performance Extreme Computing (HPEC) Conference*, Waltham, MA, September 13-15, 2016.

[Reuther 2018] A. Reuther, C. Byun, W. Arcand, D. Bestor, B. Bergeron, M. Hubbell, M. Jones, P. Michaleas, A. Prout, A. Rosa,. J. Kepner, "Scalabile System Scheduling for HPC and Big Data," *Journal of Parallel and Distributed Computing (JPDC),* vol. 111, pp. 76-92, January 2018.

[Slapnicar 2001] P. Slapničar, U. Seitz, A. Bode, and I. Zoraja, "Resource Management in Message Passing Environments," *Journal of Computing and Information Technology (CIT),* vol. 9, no. 1, 2001.

[Van Rossum 2007] G. Van Rossum, "Python Programming Language," USENIX Annual Technical Conference, 2007.

[Vavilapalli 2013] V. K. Vavilapalli, A. C. Murthy, C. Douglas, S. Agarwal, M. Konar, R. Evans, T. Gravnes, J. Lowe, H. Shah, S. Seth, and B. Saha, "Apache Hadoop YARN: Yet Another Resource Negotiator," *Proceedings of the 4th annual Symposium on Cloud Computing*, ACM, October 2013.

[Yoo 2003] A. B. Yoo, M. A. Jette, and M. Grondona, "Slurm: Simple Linux Utility for Resource Management," *Job Scheduling Strategies for Parallel Processing,* pp. 44-60, Springer Berlin Heidelberg, June 2003.

[Zhou 1993] S. Zhou, X. Zheng, J. Wang, and P. Delisle, "Utopia: A Load Sharing Facility for Large, Heterogeneous Distributed Computer Systems," *Software: Practice and Experience*, vol. 23, no. 12, pp. 1305-1336, 1993.